\documentclass[12pt]{article} 
\usepackage{graphicx}
\usepackage{epsfig} 
\usepackage{latexsym} 
\usepackage{latexsym}
\usepackage{amssymb} 
\usepackage{amsmath} 
\usepackage{cite}
\newcommand{\labell}[1]{\label{#1}}
\newcommand{\reef}[1]{(\ref{#1})}

\def\ie{{\it i.e.}}

\DeclareSymbolFont{AMSb}{U}{msb}{m}{n}
\DeclareMathSymbol{\IN}{\mathbin}{AMSb}{"4E}
\DeclareMathSymbol{\IZ}{\mathbin}{AMSb}{"5A}
\DeclareMathSymbol{\IR}{\mathbin}{AMSb}{"52}
\DeclareMathSymbol{\Q}{\mathbin}{AMSb}{"51}
\DeclareMathSymbol{\II}{\mathbin}{AMSb}{"49}
\DeclareMathSymbol{\IC}{\mathbin}{AMSb}{"43}
\DeclareMathSymbol{\IP}{\mathbin}{AMSb}{"50}
\DeclareMathSymbol{\IH}{\mathbin}{AMSb}{"48}
\DeclareMathSymbol\IA{\mathalpha}{AMSb}{"41}
\DeclareMathSymbol\IS{\mathalpha}{AMSb}{"53}

\def\Q{{\cal Q}}

\begin{document}

\begin{flushright}
USC-04-06
\end{flushright}
{\flushleft\vskip-0.9cm
\vbox{DCPT--04/27}}

\bigskip
\begin{center} 

\bigskip
\bigskip
\bigskip



{\Large\bf Unoriented Minimal Type 0 Strings}

\end{center}

\bigskip \bigskip \bigskip

\centerline{\bf James E. Carlisle${}^\sharp$ and Clifford V. Johnson${}^\natural$
}

\bigskip
\bigskip
\bigskip

\centerline{\it ${}^\sharp$Centre for Particle Theory}
  \centerline{\it Department of Mathematical Sciences }
\centerline{\it University of Durham}
\centerline{\it Durham DH1 3LE, England, U.K.}
\centerline{\small \tt j.e.carlisle@durham.ac.uk}

\bigskip
\bigskip

  \centerline{\it ${}^\natural$Department of Physics and Astronomy }
\centerline{\it University of
Southern California}
\centerline{\it Los Angeles, CA 90089-0484, U.S.A.}
\centerline{\small \tt johnson1@usc.edu}
\bigskip
\bigskip

\bigskip
\bigskip


\begin{abstract}
  
  We define a family of string equations with perturbative expansions
  that admit an interpretation as an unoriented minimal string theory
  with background D--branes and R--R fluxes.  The theory also has a
  well--defined non--perturbative sector and we expect it to have a
  continuum interpretation as an orientifold projection of the
  non--critical type~0A string for ${\hat c}=0$, the $(2,4)$ model.
  There is a second perturbative region which is consistent with an
  interpretation in terms of background R--R fluxes. We identify a
  natural parameter in the formulation that we speculate may have an
  interpretation as characterizing the contribution of a new type of
  background D--brane. There is a non--perturbative map to a family of
  string equations which we expect to be the $\hat{c}=0$ type~0B
  string. The map exchanges D--branes and R--R fluxes. We present the
  general structure of the string equations for the $(2,4k)$ type~0A
  models.

\end{abstract}
 \baselineskip=18pt \setcounter{footnote}{0}

\newpage

In what follows, we define a family of theories that have
perturbative expansions that admit interpretations as unoriented
type~0A string in the $(2,4k)$ superconformal minimal model
backgrounds, with D--branes and R--R fluxes. The definitions have a
well--defined non--perturbative regime. For the case of ${\hat c}=0$, pure
supergravity (the $(2,4)$ model) we will see that there
is a map to a system which has already been found in the literature,
arising from studying self--dual unitary matrix models in the double
scaling limit\cite{Myers:1990bb}. We argue that it is natural to
identify this system as an unorientable type~0B string, the pure
supergravity case for that theory.

The string equation for the $k=2$ member of the bosonic minimal string
theory for the $(2,2k-1)$ series can be written
as\cite{Brezin:1990rb,Douglas:1990ve,Gross:1990vs}:
\begin{equation}
  \labell{eq:PainleveI}
  -\frac13 f^{''}+f^2=z\ ,
\end{equation}
where a prime denotes $\nu\partial/\partial z$. This is the Painlev\'e
I equation.  Expanding about the large positive $z$ regime gives a
well defined perturbation theory representing an oriented bosonic
string theory that is in fact pure gravity, \ie\ the Liouville theory,
since $c=0$ for the matter theory.  The free energy is related to $f$
as $F^{''}=-f$. This model can be realized as the double scaling limit
of a one Hermitian matrix model.

In refs.\cite{Brezin:1990xr,Harris:1990kc,Brezin:1990dk,Myers:1990pp},
symmetric matrix models (orthogonal and symplectic) were studied in
the double scaling limit, and these gave rise to continuum models with
contributions from non--orientable world sheets.  The physics is then
encoded in two functions. The oriented contributions come from $f$ as
before, which (for the $k=2$ case) satisfies
equation~\reef{eq:PainleveI}, and the unoriented contributions come
from a new function, $g$, which satisfies (for this $k=2$ case) the
equation\footnote{Note that we have taken the liberty of changing the
  sign of $f(z)$ from the conventions used in
  \cite{Brezin:1990xr,Brezin:1990dk}.  This is to make contact with
  the more widespread convention in the literature.}:
\begin{equation}
  \labell{eq:geqn}
  g^3+6gg^{'}+4g^{''}-6gf-6f^{'}=0\ .
\end{equation}
Given the solution of \reef{eq:PainleveI}, equation \reef{eq:geqn}
yields three possible solutions\footnote{The meanings of which will be
  discussed later in this paper.} for the asymptotic expansion of
$g(z)$. We have:
\begin{eqnarray}
f (z) &=&\sqrt {z}-{\frac {{\nu}^{2}}{24 {z}^{2}}}-
{\frac {49 {\nu}^{4}}{1152 {z}^{9/2}}} + \cdots \nonumber \\
g_1 (z) &=& \pm \sqrt {6} z^{1/4}-{\frac {\nu}{2z}} 
\mp {\frac {5 {\nu}^{2}\sqrt {6}}{24 {z}^{9/4}}} + \cdots \nonumber \\
g_2 (z) &=&-{\frac {\nu}{2 z}}-{\frac {25 {\nu}^{3}}{24 {z}^{7/2}}}-{\frac {15745 {\nu}^{5}}{1152 {z}^{6
}}} + \cdots   \labell{eq:asympt}\
\end{eqnarray}
and the free energy of the model is given as the sum of an oriented
contribution and an unoriented one:
\begin{equation}
  \labell{eq:freeenergy}
  F=F_{o}+F_{u}\ ,\quad {\rm where}\quad F_{\mathrm{o}}^{\prime\prime} = -\frac{1}{2} f \: , \qquad F_{\mathrm{u}}^{\prime} = -\frac{1}{2} g\ . 
\end{equation}
Half of the oriented theory's free energy makes up the oriented
contribution to the free energy of the model. The remainder comes from
$g$. Unitary matrix models also have critical points, and in a double
scaling limit performed in ref.\cite{Periwal:1990gf,Periwal:1990qb}
for the simplest critical point, the following equation can be
derived:
\begin{equation}
  \labell{eq:PainleveII}
  h^3-h^{''}-hz=0\ ,
\end{equation}
which is the Painlev\'e II equation. For large positive $z$, $h$ has an
expansion with an interpretation in terms of oriented world sheets. In
fact, this model has recently been understood to be pure supergravity
for the type~0B string theory\cite{Klebanov:2003wg}, the $k=1$ member
of the superconformal series $(2,4k)$.

In ref.\cite{Myers:1990bb} self--dual unitary matrix models were
studied. It was found that there are again two contributions to
the partition function, one coming from orientable surfaces, $h$,
which satisfies equation~\reef{eq:PainleveII}; and the other,
unorientable contribution, from a function, $w$, which satisfies
the Riccati--type equation\footnote{We have changed the conventions
  of ref.\cite{Myers:1990bb} by rescaling their $t$ to $z/2$,
  their $w$ to $w/2$, and their $h^2$ to $2h^2$.}:
\begin{equation}
  \labell{eq:ggeq}
  w^{'}= \frac14 w^2 + z-\frac32 h^2  \ .
\end{equation}
Again, the free energy splits into two parts:
\begin{equation}
  \labell{eq:freeenergyU}
  F=F_{o}+F_{u}\ ,\quad {\rm where}\quad F_{\mathrm{o}}^{\prime\prime} = -\frac{1}{2} h^2 \: , \qquad F_{\mathrm{u}}^{\prime} = \frac{1}{2} w\ . 
\end{equation}

It is natural to suppose (and so we propose that it is the case) that
this is an unoriented string theory based upon some projection of the
0B theory for ${\hat c}=0$. Half of the free energy of the oriented
theory makes up the oriented sector of this new theory, and the rest
is made up of unoriented contributions.

As is by now very familiar, the physics of the even $k$ minimal
bosonic string models (of which equation~\reef{eq:PainleveI} is the
case $k=2$) suffers from a non--perturbative instability, as can be
seen from the well--known behaviour of solutions of the equation for
$f$.  The unoriented theory simply inherits this behaviour.

It is known that there is a family of models that have
the same perturbative behaviour as the string equations for the
bosonic minimal models; but have better non--perturbative behaviour,
defining a family of non--perturbatively stable oriented minimal
string theories with the same perturbative content. An example
equation is:
\begin{equation}
f{\cal R}^2-\frac{1}{2}{\cal R}{\cal R}^{''}+\frac{1}{4}({\cal R}^{'})^2
  =\nu^2\Gamma^2\ .\labell{eq:nonpert}
\end{equation}
where 
\begin{equation}
  \labell{eq:defR}
  {\cal R}\equiv   -\frac13 f^{''}+f^2-z\ ,
\end{equation}
and for $\Gamma=0$ they have the same large positive $z$ expansion as
equation~\reef{eq:PainleveI}. This follows from the fact that ${\cal
  R}=0$ is a solution to the equation when $\Gamma=0$. There are
solutions that solve the equation~\reef{eq:nonpert} which are not
solutions of ${\cal R}=0$ however, but nevertheless have the same
large positive $z$ asymptotic expansions as the solutions of ${\cal
  R}=0$. These solutions define theories with well--defined
non--perturbative physics\cite{Dalley:1992qg,Dalley:1992vr}; and they have now been identified\cite{Klebanov:2003wg} as the type~0A
minimal string theories, for the $(2,4k)$ series. The case we have
given above is in fact $k=2$.  The string equation above captures more
than that however\cite{Dalley:1992br}. For $\Gamma\neq0$ it includes
a new feature. For large positive $z$, the parameter $\Gamma$
corresponds to there being $\Gamma$ D--branes in the
background\cite{Dalley:1992br}, while for large negative $z$ it
corresponds to $\Gamma$ half--units of R--R
flux\cite{Klebanov:2003wg}.

The simplest model is in fact the choice $k=1$, which is pure type 0A
world--sheet supergravity, and has:
\begin{equation}
  \labell{eq:kone}
{\cal R}\equiv f-z\ ,  
\end{equation}
In fact, the case $k=1$ is very special. There is a
change of variables\cite{Morris:1991cq} which takes the
equation~\reef{eq:nonpert} (with ${\cal R}$ given in
equation~\reef{eq:kone}, and with $\Gamma=0$ (for now)) and maps it to
the equation~\reef{eq:PainleveII}. The map is:
\begin{equation}
  \labell{eq:timsmap}
  f=h^2+z\ ,
\end{equation}
together with the exchange $z\to -z$. This means, as first pointed out
in ref.\cite{Klebanov:2003wg}, that the physics of pure 0A
supergravity and pure 0B supergravity are non--perturbatively related.
Non--perturbative because these systems each have two distinct
asymptotic regimes, large positive $z$ and large negative $z$ which
are separated by a strongly coupled regime. The map above exchanges
these regimes\footnote{As pointed out in ref.\cite{Johnson:2004ut},
  this relation may be an important example. The large positive $z$
  physics of the 0A model is identical to that of the $k=1$ {\it
    bosonic} minimal string model, which is a topological string
  theory. Pure 0A can therefore be thought of as a non--perturbative
  completion of the topological theory. Moreover, this completion can be
  interpreted as a 0B model.  It is a strong--weak coupling duality
  for topological strings, which deserves further investigation.}.
More generally, for non--zero $\Gamma$, the map in
equation~\reef{eq:timsmap}, together with $z\to-z$, takes the 0A string
equation~\reef{eq:nonpert} to a generalization\cite{Klebanov:2003wg}
of the string equation~\reef{eq:PainleveII}:
\begin{equation}
  \labell{eq:PainleveIImore}
   h^3 - h^{''} - h z - \frac{\nu^2 \Gamma^2}{h^3} = 0 \ ,
\end{equation}
which, for the type~0B model now, has an interpretation in terms of
having $\Gamma$ background D--branes in one perturbative regime and
$\Gamma$ units of flux in the opposite perturbative regime.

A natural question to ask is whether these 0A equations have a
generalization to include an interpretation of the presence of {\it
  non--orientable} surfaces in (at least one of) the perturbative
regimes. We shall now show that there is a very natural generalization to
include this interpretation.

First, we digress a short while to introduce an elegant differential
operator language in which the non--orientable world--sheet physics of
the bosonic matrix models can be cast\cite{Brezin:1990dk}. The basic
object in terms of which everything follows is the operator:
\begin{equation}
  \labell{eq:Qoperator}
  Q=d^2-f(z)\ ,\quad {\rm where}\quad  d\equiv \nu {\partial\over \partial z}\ .
\end{equation}
For the $k$th bosonic model, one forms the following truncation of a
fractional power of $Q$, which we shall denote\cite{Douglas:1990dd} as
$P$:
\begin{equation}
  \labell{eq:Poperator}
  P=Q^{k-\frac12}_+\ .
\end{equation}
The fractional power is defined by requiring that there exist
an operator $Q^{\frac12}$ such that it squares to give $Q$. In doing
this, one needs a definition of $d^{-1}$, and this is such
that\cite{Gelfand:1975rn,Drinfeld:1984qv}:
\begin{equation}
  \labell{eq:dminusone}
  d^{-1}f= fd^{-1}+\sum_{i=1}^{\infty}(-1)^i f^{(i)}d^i\ ,
\end{equation}
where $f^{(i)}$ means the $i$th $z$--derivative of $f$.  The plus sign subscript in the equation \reef{eq:Poperator} above means that we take only
positive powers of $d$. The string equations are formulated as the
realization of the fundamental relation\cite{Douglas:1990dd}:
\begin{equation}
  \labell{eq:pqone}
  [P,Q]=1\ ,
\end{equation}
where one integrates once with respect to $z$ to get the final string
equation. In the case of $k=2$ we have:
\begin{equation}
  \labell{eq:Pktwo}
  P=d^3-\frac{3}{4}\left\{d,f\right\}\ ,
\end{equation}
which leads to equation~\reef{eq:PainleveI}.  The unoriented
sector is defined in terms of the above objects as follows.  Require
that $P$ be factorizable into the following form:
\begin{equation}
  \labell{eq:Pfactortwo}
  P=(d^2+\frac{1}{2}gd+h)(d-\frac{1}{2}g)\ ,
\end{equation}
leading to two equations, which when combined give
equation~\reef{eq:geqn}.  We point out here that
equation~\reef{eq:geqn} can be written as
\begin{equation}
   \labell{eq:Tee2}
  T^{'}+gT=0\ ,
\end{equation}
where 
\begin{equation}
  \labell{eq:Tee}
  T\equiv g^{'}+\frac{1}{4}g^2-\frac{3}{2}f\ .
\end{equation}
A natural sub--family of solutions are those which satisfy $T=0$,
which is the following Riccati--type equation:
\begin{equation}
  \labell{eq:Riccati}
  f=\frac{1}{6}g^2 +\frac{2}{3}g^{'}\ .
\end{equation}
In fact, as pointed out in ref.\cite{Brezin:1990xr}, studying the
equation~\reef{eq:Riccati} is equivalent to the more restrictive
factorization of $P$:
\begin{equation}
  \labell{eq:factormore}
  P=(d+\frac12 g)d(d-\frac12 g)\ .
\end{equation}
It turns out that the $g_1(z)$ perturbative expansions from
equation~\reef{eq:asympt} are the solutions to the $T\!\!=\!0$
equation; whereas $g_2(z)$ corresponds to a solution that satisfies
equation~\reef{eq:Tee2}, but not $T\!\!=\!0$. It is actually very
interesting that there are these three different solutions for $g(z)$.
In refs.\cite{Brezin:1990xr,Harris:1990kc}, one of the $g_1$ solutions
is associated with the orthogonal matrix ensemble; the other is
associated with the symplectic ensemble. Although the third solution,
$g_2$, is a solution of an equation derived from the matrix model, it
has yet to receive an interpretation in terms of a particular
ensemble. However, there is good cause to expect that it may also
correspond to viable physics.

The same operator language that was used above in the $[P, Q]=1$
bosonic case can also be used to formulate the type~0A models. This is
because, as shown in ref.\cite{Dalley:1992qg,Dalley:1992vr}, the
differential operator structure which underlies the minimal bosonic
string equations can be used to define the minimal type~0A string
equations as well.  We then apply a factorization condition to the
appropriate operator involved in this new definition. It works as
follows:

The first derivative of the string equation arises by defining, for a
particular $k$, the operator\cite{Dalley:1992qg}:
\begin{eqnarray} 
\labell{eq:PToperator}
\tilde{P} = Q^{k+\frac{1}{2}}_+ - \frac{1}{2} z d\ , 
\end{eqnarray}
and imposing the fundamental equation:
\begin{equation}
  \labell{eq:PTQQ}
  [\tilde{P}, Q] = Q\ .
\end{equation}
Indeed, this is an equation stating scale invariance, in contrast to
the previous one (equation~\reef{eq:pqone}) which states translation
invariance. This fits with the fact that the scaled eigenvalue
distribution in the underlying matrix model is defined on $\IR^+$ in
one case, and $\IR$ on the other\cite{Dalley:1992vr}.
Equation~\reef{eq:PTQQ} then gives the first derivative of the string
equation~\reef{eq:nonpert}, and $\nu^2\Gamma^2$ arises as an
integration constant\footnote{There are many other ways to derive
  equation~\reef{eq:nonpert}. See
  refs.\cite{Dalley:1992qg,Dalley:1992vr} for matrix model
  derivations}.

Now we propose that it is natural to define the unoriented
contribution by requiring that $\tilde{P}$ factorizes in the same way
as $P$. 
This will define physics for all the different $k$ cases in a manner
analogous to the multicritical points of ref.\cite{Brezin:1990dk},
(which we propose to be the type~0A $(2,4k)$ unoriented models) but
let us work with the case of $k=1$ (pure supergravity) to see how
things work explicitly.

Before proceeding, we note that the scaling operator defined in
equation~\reef{eq:PToperator} is more restrictive than necessary, and
misses potentially important physics. More generally, adding a
constant, $\nu C_0$ to the operator $\tilde{P}$ gives an operator
which is just as good as a scaling operator, the
equation~\reef{eq:PTQQ} remaining true.

So we have\footnote{Note that we have chosen the negative sign of the
  square root of $Q$ here to make things appear more natural
  later.}:
\begin{eqnarray}
\tilde{P} = -d^3 + \frac{3}{2} f d + \frac{3}{4} f^\prime - \frac{1}{2} z d + \nu C_0 = -(d^2 + \frac{1}{2}g d + h)(d - \frac{1}{2}g)\ , 
\end{eqnarray}
 and upon  expanding and equating powers of derivatives we find two equations:
\begin{eqnarray}
-g^\prime +h - \frac{1}{4} g^2 = -\frac{3}{2}f +\frac{1}{2} z \, , \qquad \frac{1}{2} h g + \frac{1}{4} g g^\prime + \frac{1}{2} g^{\prime \prime} = \frac{3}{4} f^\prime + \nu C_0\ , 
\end{eqnarray}
from which, after elimination of $h$ we obtain:
\begin{eqnarray} \labell{eq:F1}
g^3 + 6 g g^\prime + 4 g^{\prime \prime} - 6 g f - 6 f^\prime + 2 z g = 8 \nu C_0\ . 
\end{eqnarray}
Given the experience of the previous two cases, the question arises as
to whether there is a first order Riccati--type equation to which this
is equivalent. Noting that ref.\cite{Brezin:1990dk} showed that one
can restrict to that form by a more specific factorization of $P$, we
try the same thing for~${\tilde P}$:
\begin{eqnarray}
\tilde{P} = -d^3 + \frac{3}{2} f d + \frac{3}{4} f^\prime - \frac{1}{2} z d + \nu C_0 = -(d + \frac{1}{2}g) d (d - \frac{1}{2}g) \ ,
\end{eqnarray}
giving:
\begin{eqnarray}
-\frac{1}{4} g^2 - g^\prime = -\frac{3}{2} f + \frac{1}{2} z \, , \qquad \frac{1}{4} g g^\prime + \frac{1}{2}g^{\prime \prime}  = \frac{3}{4} f^\prime + \nu C_0\end{eqnarray}
In order for these two equations to make sense we require that they
must have the same content. To do this we must set $C_0 = -1/4$, which means the equation for $g$ can be written simply as:
\begin{equation}
  \labell{eq:Sequation}
  S=0\ ,
\end{equation}
where 
\begin{equation}
  \labell{eq:Sdef}
  S =g^{'}+\frac14 g^2-\frac32 f +\frac12 z\equiv T+\frac12 z\ .
\end{equation}
Using this notation we find that the result of the less restrictive factorization \reef{eq:F1} can be expressed as (again with $C_0 = -1/4$): 
\begin{equation}
  \labell{eq:Sequation2}
  S^{'} + g S =0\ ,
\end{equation}

This is very pleasant indeed, since using the map~\reef{eq:timsmap},
together with $z\to-z$ and the identification $g=w$, the equation
$S=0$ becomes equation~\reef{eq:ggeq} derived for the symmetric
unitary matrix models, which as we said earlier pertain to the type~0B
system!  So we have extended the map between 0A and 0B to one between
their unoriented descendants. This also serves to put our computations
on a firm double scaled matrix model footing.

It is interesting to examine the behaviour of the solutions in each
asymptotic regime. We have solutions of equation~\reef{eq:nonpert} for
$f(z)$ in the large positive $z$ and negative $z$ directions,
respectively:
\begin{eqnarray}
f_+(z) &=&  z+{\frac {\nu\,\Gamma}{\sqrt {z}}}-{\frac {{\nu}^{2}{\Gamma}^{2}}{2 {z}^{2}}}+{
\frac {5 {\nu}^{3}\Gamma\, \left( 4\,{\Gamma}^{2}+1 \right) }{
32 {z}^{7/2}}}-{\frac {{\nu}^{4}{\Gamma}^{2} \left( 8\,{\Gamma}^{2}+7 \right) }{8 {z}^{5}}}
 + \cdots \nonumber \\
f_-(z) &=& {\frac {{\nu}^{2} \left( 4\,{\Gamma}^{2}-1 \right) }{4 {z}^{2}}}\left(1+{\frac { 4\nu^2  \left( 4\,{\Gamma}^{2}-9 \right) }{{z}^{3}}}+{\frac {7 {\nu}^{4}  
 \left( 4\,{\Gamma}^{2}-9 \right)  \left( 4\,{\Gamma}^{2}-21
 \right) }{16 {z}^{6}}}
 + \cdots\right)\ .
 \labell{eq:Orient}
\end{eqnarray}
Looking in the positive $z$ expansion, for example, integrating twice
and dividing by $\nu^2$ reveals an expansion in the dimensionless
parameter $g_s=\nu/z^{3/2}$. The first term apparently corresponds to a
sphere term, but it is non--universal and should be dropped, while the
next two correspond to the contributions of the disc and annulus,
respectively.  The relevant diagrams come with a factor
$g_s^{2h+b-2}\Gamma^b$, where $b$ is the number of boundaries and $h$
is the number of handles. In the negative $z$ expansion, $\Gamma^2$ is
to be taken as representing a single R--R flux
insertion\cite{Klebanov:2003wg}, and so the first term is the torus,
(the $\Gamma$ independent term) and sphere with a flux insertion.

\begin{figure}[ht]
\begin{center}
  \includegraphics[scale=0.6]{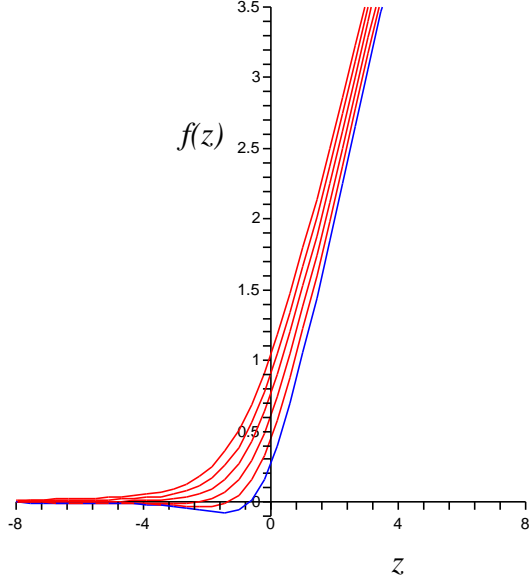}
\end{center}
\caption{\small The non--perturbative solution for the function,
  $f(z)$, which contributes to the oriented sector of the model, for a
  range of positive values of $\Gamma$. The bottom curve is the
  $\Gamma=0$ case.}
\label{fig:gammaplots_f}
\end{figure}

There are unique non--perturbative completions of these solutions, and
some of these are plotted in figure~\ref{fig:gammaplots_f} for a range
of values of $\Gamma$.
With each of these asymptotics comes two choices for $g(z)$, as
solutions of $S=0$. For the large positive and negative $z$
directions, we have the choices, respectively:
\begin{eqnarray}
g_{+,1}(z) &=& \pm 2\,\sqrt {z}+{\frac {\nu\, \left( \pm 3\,\Gamma-2 \right) }{2 z}}-
{\frac {{\nu}^{2} \left( \pm 21\,{\Gamma}^{2}-36\,\Gamma \pm 20 \right) }{16{z}^{5/2}}}
 + \cdots \nonumber \\ 
g_{-,1}(z) &=& \pm i\sqrt {2}\sqrt {z}-{\frac {\nu}{z}} \mp {\frac {i\sqrt {2}{\nu}^{2}
 \left( 12\,{\Gamma}^{2}-13 \right) }{8 {z}^{5/2}}}-{\frac {3 {
\nu}^{3} \left( 12\,{\Gamma}^{2}-13 \right) }{4 {z}^{4}}}
+ \cdots
\labell{eq:gexp1}
\end{eqnarray}
and we can also have third and fourth solutions that satisfy
(\ref{eq:Sequation2}) but not (\ref{eq:Sequation}):
\begin{eqnarray}
g_{+,2}(z) &=& -{\frac {\nu}{z}}+{\frac {9 {\nu}^{2}\Gamma}{4 {z}^{5/2}}}-{\frac {15 {\nu}^{3} \left( 3\,{\Gamma}^{2}+2
 \right) }{8 {z}^{4}}}
 + \cdots \nonumber \\
g_{-,2}(z) &=& -{\frac {\nu}{z}}-{\frac {3 {\nu}^{3} \left( 12\,{\Gamma}^{2}-
13 \right) }{4 {z}^{4}}}-{\frac {3 {\nu}^{5} \left( 336\,{\Gamma}^{4}-3384\,{\Gamma}^{2}+3085 \right) }{16 {z}^{7}}}
 + \cdots
\labell{eq:gexp2}
\end{eqnarray}
An examination of the positive $z$ regime for the first case
($g_{+,1}$) shows (after integrating once and dividing by $\nu$) an
expansion in $g_s$ again.  It is clear that the first two terms are
the contributions of $\IR {\rm P}^2$, the M\"obius strip, and the Klein
bottle, respectively. The diagrams come with a factor
$g_s^{2h+b+c-2}\Gamma^b$ where $c$ is the number of crosscaps. In the
negative $z$ regime we have an interpretation with just R--R fluxes
again, the first term being $\IR {\rm P}^2$ and the next the Klein bottle.
The next has contributions from closed surfaces with the following:
$c=1, h=0$ and a flux insertion; $c=3, h=0$; $c=1, h=1$.

For the second case, there is again a surface interpretation
consistent with there being D--branes in the positive $z$ regime, and
fluxes in the negative $z$ regime. However, there are missing orders
that deserve some explanation, which is perhaps to be found in a study
of the continuum theory\footnote{There have been some recent studies
  of non--critical string theory on unoriented surfaces which may be
  relevant\cite{Hikida:2002bt,Nakayama:2003ep,Gomis:2003vi,Bergman:2003yp}.}.

\begin{figure}[ht]
\begin{center}
  \includegraphics[scale=0.6]{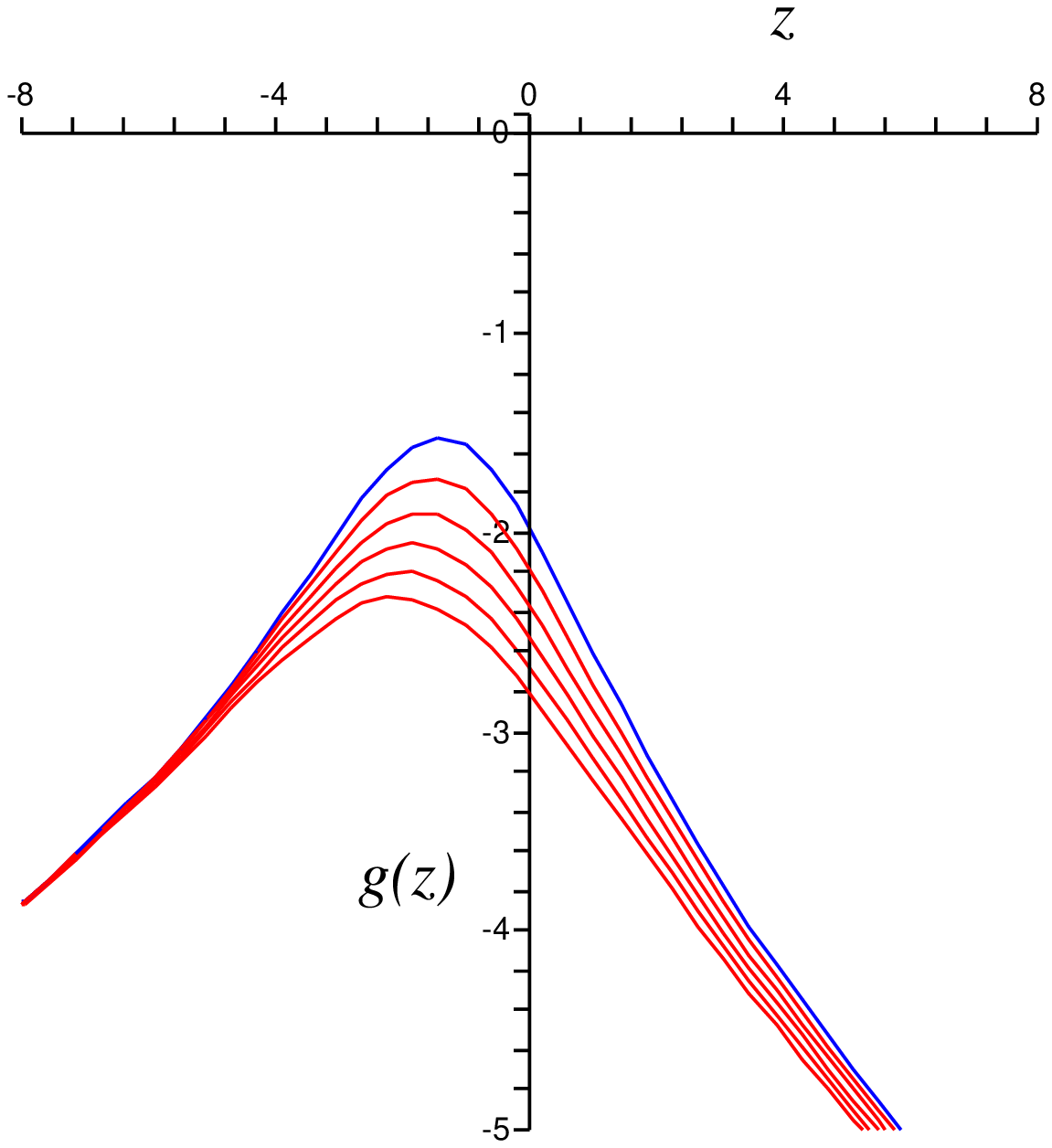}\includegraphics[scale=0.6]{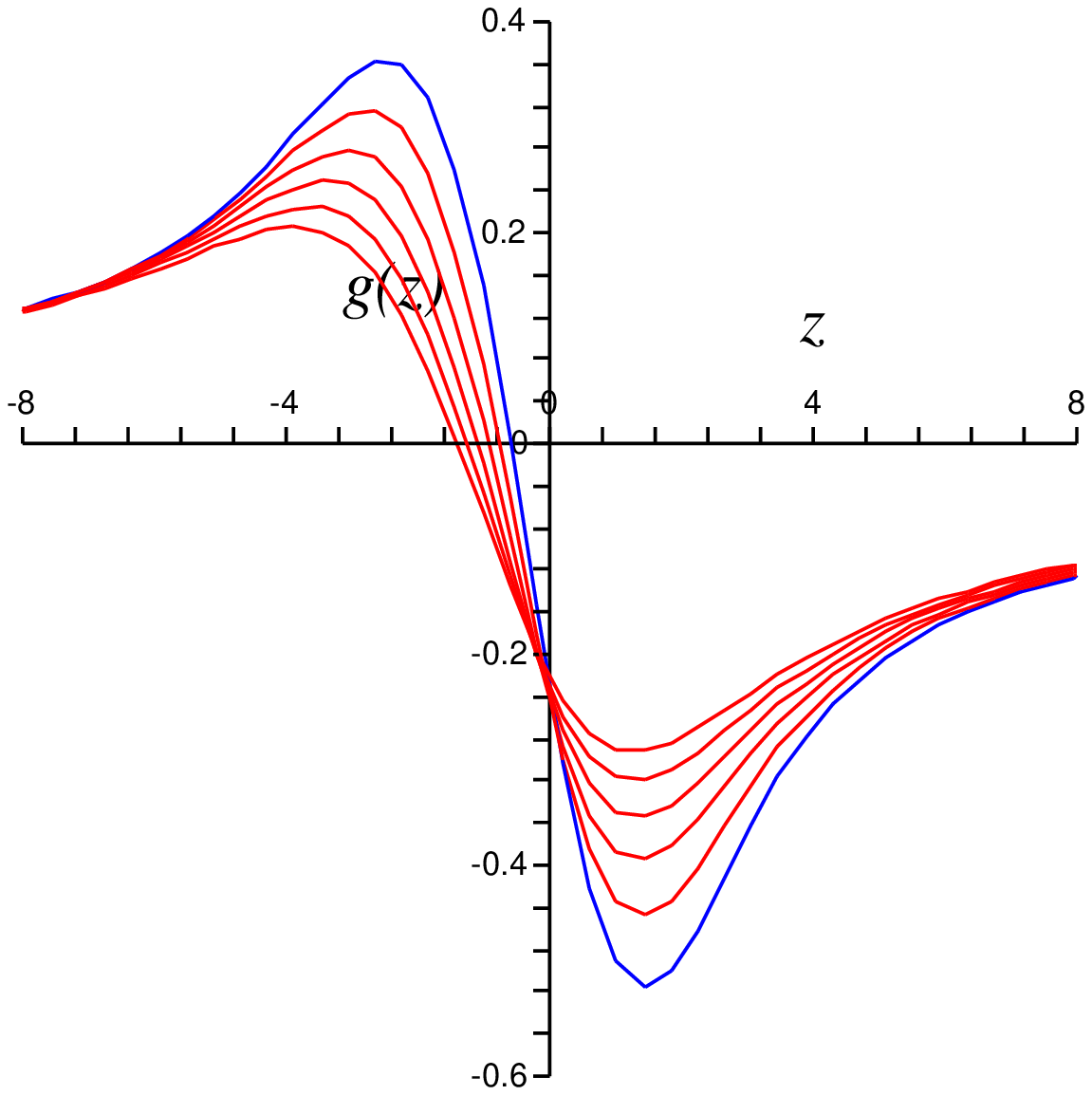}
\end{center}
\caption{\small Two non--perturbative solutions for the function
  $g(z)$ which contributes to the unoriented sector of the model, for
  a range of positive values of $\Gamma$. In both cases the top curve
  is the $\Gamma=0$ case.}
\label{fig:gammaplots_ga}
\end{figure}

Plots of the non--perturbative completions of these two cases are
given in figure~\ref{fig:gammaplots_ga}.
As in the case of the $[P,Q]=1$ formulation, it is tempting to
speculate that all of these perturbative expansions for $g(z)$ are on
equal footing; that is, they are equally valid in non-orientable
string theory. Whether or not all these models stem from matrix models
is also a question worth investigating.

Having arrived at the case $C_0=-1/4$, it is natural to wonder whether
the other choices for $C_0$ are physical as well.  After a little
algebra, we find that we can rewrite equation~\reef{eq:F1} as the
following:
 \begin{equation}
  \labell{eq:Sequationagain}
  S^{'}+gS=2\nu\left(C_0+\frac14\right) \equiv \nu G\ ,
\end{equation}
where we have defined the constant $G$ for later convenience. The case
before was $G=0$. To what do other values of $G$ correspond? The case
of $C_0=0$ ($G = 1/2$), which corresponds to the $\tilde{P}$ operator
originally defined without the shift, gives the following expansions
for $g$ (again two choices in each regime). For large positive $z$ we
have:
\begin{eqnarray} \labell{eq:Unorient}
g_{+, 1}(z) &=& \pm 2\,\sqrt {z}+{\frac {3 \nu\, \left( \pm 2\,\Gamma-1 \right) }{4 z}}-
{\frac {{\nu}^{2} \left( \pm 84\,{\Gamma}^{2}-120\,\Gamma \pm 47 \right) }{64 {z}^{5/2}}}
 + \cdots \nonumber\\
g_{+, 2}(z) &=& -{\frac {3 \nu}{2 z}}+{\frac {3 {\nu}^{2}\Gamma}{{z}^{5/2}}}-{\frac {3 {\nu}^{3} \left( 76\,{\Gamma}^{2}+77
 \right) }{32 {z}^{4}}}
 + \cdots 
\end{eqnarray}
while for the large negative $z$ regime we have: 
\begin{eqnarray}
g_{-, 1}(z) &=& \pm i\sqrt {2}\sqrt {z}-{\frac {3 \nu}{2 z}} \mp {\frac {i\sqrt {2}{\nu
}^{2} \left( 24\,{\Gamma}^{2}-47 \right) }{16 {z}^{5/2}}}-{\frac {21 {\nu}^{3} \left( 8\,{\Gamma}^{2}-17 \right) }{16 {z}
^{4}}}
+ \cdots\nonumber \\
g_{-, 2}(z) &=& -{\frac {3{\nu}^{3} \left( 4\,{\Gamma}^{2}-1 \right) }{2 {z}^{4
}}}-{\frac {6 {\nu}^{5} \left( 8\,{\Gamma}^{4}-54\,{\Gamma}
^{2}+13 \right) }{{z}^{7}}}
 + \cdots
\end{eqnarray}
Again, a surface interpretation can be given along the lines of those
discussed below equations~\reef{eq:gexp1} and~\reef{eq:gexp2}.

Leaving $G$ as a parameter in the solutions (rather analogous to
$\Gamma$, which controls the number of background D--branes or R--R
fluxes), these equations yield the following solutions for $g(x)$:
\begin{eqnarray}
g_{+,1}(z) &=& \pm 2\,\sqrt {z}+{\frac {\nu\, \left( \pm 3\,
\Gamma - 2+G  \right) }{2 z}}-{\frac {{\nu}^{2} \left( \pm 21\,{\Gamma}^{2}+12\Gamma(G-3) \pm 3\,{G}^{2} \mp 18\,G \pm 20
 \right) }{16 {z}^{5/2}}}
+ \cdots \nonumber\\ 
g_{-,1}(z) &=& \pm i\sqrt {2}\sqrt {z}-{\frac {\nu\, \left( G
+1 \right) }{z}} \mp {\frac {i\sqrt {2}{\nu}^{2} \left( 12\,{\Gamma}^{2}-6\,{G}^{2}-18\,G-13 \right) }{8 {z}^{5/2}}}
 + \cdots \nonumber\\
g_{+,2}(z) &=& -{\frac {\nu\, \left( G+1 \right) }{z}}+
{\frac {3 {\nu}^{2}\Gamma\, \left( 2\,G+3 \right) }{4 {z}^{5/2}}}-{\frac {{\nu}^{3} \left( 3{\Gamma}^{2}(8G+15)+2{G}^{2}(G+9)+46\,G+30 \right) }{8 {z}^{4}}}
 + \cdots \nonumber \\
g_{-,2}(z) &=& {\frac {\nu\, \left( 2\,G-1 \right) }{z}}
+{\frac {{\nu}^{3} \left( 24\,{\Gamma}^{2}G-36\,{\Gamma
}^{2}-16\,{G}^{3}+72\,{G}^{2}-98\,G+39 \right) }{4 {z}^{4}}}
 + \cdots
\end{eqnarray}
\begin{figure}[ht]
\begin{center}
  \includegraphics[scale=0.6]{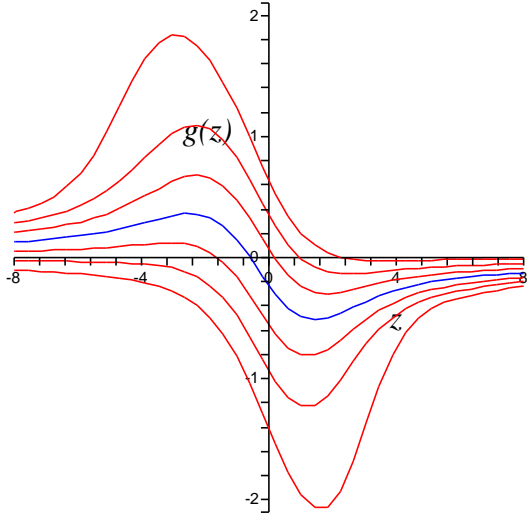}
\end{center}
\caption{\small The non--perturbative solution for the function $g(z)$
  which contributes to the unoriented sector of the model, for a range
  of values of $G$. The middle curve is the case of $G=0$.}
\label{fig:qplots_g}
\end{figure}
These solutions interpolate between those displayed earlier for the
cases $G=0$ and $G=1/2$.  Figure~\ref{fig:qplots_g} shows some sample
plots of the full non--perturbative $g(z)$ for varying values of $G$.
Upon examination of the expansions, it is tempting to interpret $G$ as
a parameter controlling some aspect of the background, which is
present only for the unoriented case.  One possibility is that it
incorporates the presence of some number, $G$, of a new type of
D--brane that is present only in non--orientable diagrams (since it
does not appear in the contributions from oriented surfaces).  A
candidate such brane could be one which is forced to remain stuck at
(the analogue of) orientifold planes, a sort of fractional brane. This
deserves further investigation. It is the case $G=0$ that has a known
matrix model interpretation (by virtue of the map to self--dual
unitary matrix models discussed earlier), and this would correspond to
having such branes absent.

A possible clue to the nature of $G$ might be found in an examination
of the instanton corrections to the expansions (terms exponentially
small for small $g_s$), which often have an interpretation as
D--branes in the theory\cite{Shenker:1990uf,McGreevy:2003kb}. Some
standard computations reveal that there are a number of instantons
associated with each solution. These have actions (D--brane tensions)
given by:
\begin{eqnarray}
  \label{eq:tensions}
f_+ :\quad \frac43\frac{1}{g_s}\ ;  \quad  
g_{+, 1} :\quad \frac43\frac{1}{g_s} 
\quad \mathrm{or} \quad \frac23\frac{1}{g_s}\ ;  \quad 
g_{+, 2} :\quad \frac23\frac{1}{g_s}\ ;   \quad 
g_+ :\quad \frac23\frac{1}{g_s}\ . 
\end{eqnarray}
In each case, the positive $z$ instanton has a negative $z$
counterpart which has an action precisely $1/\sqrt{2}$ times
smaller\footnote{It is not clear if these latter represent branes
  given the R--R flux interpretation in that perturbative regime.}.
Here, the solution $g_+$
is the one obtained directly from the equation $S=0$.  In each
perturbative regime it appears that the tension of the unoriented
expansion instantons takes one of two values: the same value
associated to the oriented instanton; and half the value associated to
the oriented instanton. It is therefore possible that this second
unoriented instanton is the extra D--brane that is in the background,
whose number is measured by $G$. This needs further work to properly
establish however; but our picture is suggestive, if somewhat
speculative.

The functions which give rise to $f$ and $g$ in the matrix model arise
from orthogonal polynomial coefficients that come from effectively
splitting the eigenvalues $\lambda_i$ of the matrices into the ones
with even labels ($\lambda_{2j}$) and the ones with odd labels
($\lambda_{2j+1}$), respectively. The physics that arises from
quantities controlled by $f$ are known to have an interpretation in
terms of an eigenvalue distribution, the properties of which in the
scaling limit can be interpreted in terms of background
D--branes. In fact there is a direct map between the function
$f$ and the eigenvalue distribution which can be studied on the sphere
using an integral representation\cite{Bessis:1980ss}, or in terms of the
resolvent of the operator $Q$\cite{David:1990ge}. The latter is
particularly useful for going beyond the sphere and incorporating the
effects of $\Gamma$ in the background, as recently shown in
ref.\cite{Johnson:2004ut}.  It is natural to wonder if there is an
effective eigenvalue distribution associated to the function $g$ that
controls the physics of the unoriented sector. The instantons we
computed above would have a natural home there in terms of the value
of an effective potential at an extremum, and may correspond to a new
type of D--brane. It seems likely (since $g$ arises as part of the
operator $\tilde{P}$ which is made of a fractional power of $Q$) that
there is some natural geometrical information to be extracted from $g$
analogous to an eigenvalue distribution and its effective potential.
This is worth exploring.

Finally, it is known\cite{Johnson:1994vk,Johnson:2004ut} that the
introduction of $\Gamma$ D--branes (or R--R flux units) into a closed
minimal string background can be performed very elegantly in the
language of the underlying twisted boson ($\varphi$, which is related
to the loop operator and the resolvent) by simply acting with a vertex
operator $V=:\exp(-\Gamma\varphi/\sqrt{2}):$ It would be interesting to
see how the unoriented theory is described in this framework. Perhaps
the role of $G$ may be best uncovered in this context.

{\bf Acknowlegements:} JEC is supported by an EPSRC studentship, and
also thanks the University of Southern California's Department of
Physics and Astronomy for hospitality.

\providecommand{\href}[2]{#2}\begingroup\raggedright\endgroup

\end{document}